\documentclass{article}
\usepackage[utf8]{inputenc}
\usepackage{amsmath,amssymb,amsfonts,amsthm,amscd}
\usepackage{graphicx}
\usepackage{color}
\usepackage{comment}
\usepackage[T1]{fontenc}
\usepackage{authblk}
\usepackage{caption}
\usepackage{subcaption}
\usepackage{comment}
\usepackage{pst-node}
\usepackage{tikz-cd} 
 \usepackage{color}
\newcommand{\AM}{\textcolor{black}}
\newcommand{\MC}{\textcolor{black}}

\usepackage{cancel}

\newtheorem{remark}{Remark}[section]

\newcommand{\rvline}{\hspace*{-\arraycolsep}\vline\hspace*{-\arraycolsep}}

\newcommand{\mcadded}[1]{{\color{black}#1}}

\def\w{\omega}
\def\pe{p^\varepsilon}
\def\qe{q^\varepsilon}
\def\ze{z^\varepsilon}
\def\ve{\varepsilon}
\def\ye{y^\varepsilon}

\title{A hybrid approach to model reduction of Generalized Langevin Dynamics}

\author[1]{Matteo Colangeli\thanks{matteo.colangeli1@univaq.it}}
\author[2]{Manh Hong Duong\thanks{h.duong@bham.ac.uk}}
\author[3]{Adrian Muntean\thanks{adrian.muntean@kau.se}}
\affil[1]{Department of Information Engineering, Computer Science and Mathematics,
University of L'Aquila, Italy.}
\affil[2]{School of Mathematics,
University of Birmingham,
UK.}
\affil[3]{Department of Mathematics and Computer Science \& Centre for Societal Risk Research (CSR), Karlstad University, Sweden.}

\begin{document}

\maketitle

\begin{abstract}
\AM{We consider a classical model of non-equilibrium statistical mechanics accounting for non-Markovian effects, which is referred to as the Generalized Langevin Equation in the literature. We derive reduced} Markovian descriptions obtained through the \AM{neglection} of inertial terms and/or heat bath variables. The adopted reduction scheme relies on the framework of the Invariant Manifold \AM{method, which} allows to retain the slow degrees of freedom from a multiscale dynamical system. Our approach is also rooted on the Fluctuation-Dissipation Theorem, which helps preserve the proper dissipative structure of the reduced dynamics. We highlight the appropriate time scalings introduced within our procedure, and also prove the commutativity of \AM{selected} reduction paths.
\end{abstract}

\section{Introduction}
Complex systems, \AM{commonly encountered in the study of physical and biological processes} as well as in social sciences (e.g. in \AM{molecular dynamics, motion of human crowds or fish swarms}, opinion formation in \AM{political elections}, etc.), are often described by \AM{coupled} systems of stochastic differential equations (SDEs). \AM{These} are meant to model the time evolution of a large number of microscopic constituents \mcadded{interacting with one another, possibly also coupled to some external reservoirs}. The stochastic character of the equations is introduced to account for a certain degree of uncertainty in the description of the phenomenon -- which is intrinsic in many applications, and may originate from various sources, e.g. noisy and rapidly changing environments -- as well as the presence of unavoidable human estimate errors \cite{ bianconi2023complex}. The noise term thus allows to meaningfully encode the details of the microscopic interactions with certain unknown factors (unresolved particles or reservoirs) endowed with prescribed statistical properties. This approach turns out to be of vital importance in the analysis of stochastic climate models, see e.g. the seminal paper by K. Hasselmann \cite{hasselmann1976stochastic} and the recent review \cite{lucarini2023theoretical}.

However, the resulting SDEs are often hard to treat analytically, and even their numerical characterization may not be easily accessible, typically owing to the 
high dimensionality of the model (large number of degrees of freedom, or of the involved parameters, etc.). It is thus of both theoretical and practical \mcadded{relevance to describe large complex systems through simpler and lower dimensional ones, while still preserving the essential details of the original models.}
Over the last two decades, such research endeavour has gradually developed and systematized into the growing field of model reduction \MC{methods}, which are now applied not only to SDEs but also to other types of mathematical models, such as ODEs and PDEs \cite{givon2004extracting,gorban2006model,Ott05,Snowden,Rupe,benner2017model}.

A natural and largely studied reduction approach consists in projecting the full Markovian dynamics onto a manifold parametrized by a set of relevant, typically slow, variables (for instance, certain marginal coordinates), sometimes called ``collective variables'' in the literature \cite{hartmann}. However, within this approach, a non-Markovian dynamics, characterized by the presence of some memory terms, 
is typically obtained \cite{Zwanzig}. It is then desirable  to restore the Markovian structure of the original process, so as to avoid the onset of memory terms in the dynamics. In the set-up of fast-slow systems, which constitute the focus of most existing works in the literature, as the ratio of fast-to-slow characteristic time-scales increases, the fast dynamics equilibrates more and more rapidly with respect to the slow ones; consequently,  a fully \MC{Markovian} reduced description for the evolution of the slow variables can be derived in the limit of a perfect time-scale separation, see e.g. \cite{Zwanzig,Ghil} and the monographs \cite{pavliotis2008multiscale, gorban2006model}. More recently, there has been \AM{an} increasing  interest towards the development of techniques of model reduction not relying on the 
aforementioned separation \cite{TonyRoberts}. In \cite{Wouters2019,Wouters2019b} the authors present a reduction procedure, based on the Edgeworth expansion, for both deterministic and stochastic dynamics displaying a moderate time-scale separation. In \cite{Checkroun1,Checkroun2} the Ruelle–Pollicott resonances are investigated in a reduced state space in which time-scales are only weakly separated. 
The papers~\cite{Legoll2010,Lu2014, zhang2016effective,duong2017variational,Duong2018, Legoll2019,Lelievre2019,Hartmann2020} deal, instead, with diffusion processes for which no time-scale separation is explicitly invoked, advocating an approach \AM{based on} conditional expectations.

In this paper we apply a reduction procedure to the generalized Langevin dynamics, which is a fundamental model in nonequilibrium statistical mechanics. Our goal is essentially two-fold. On the one hand, we aim to recover and connect, within a common theoretical framework, different Langevin equations widely known and debated in the literature, describing underdamped or overdamped dynamics, and sometimes also including heat bath variables. On the other hand, we also want to highlight meaningful corrections to the foregoing equations, originating from the lack of a perfect time-scale separation.
The work is organized as follows. In Sec. \ref{sec:GLE} we introduce the Generalized Langevin Equation and discuss the various reduction paths examined throughout the manuscript. In Secs. \ref{sec:under} we study the derivation of the underdamped Langevin equation from the Generalized one, whereas in Sec. \ref{sec:over} we pursue the reduction procedure further in order to obtain the overdamped Langevin from the underdamped one. In Sec. \ref{sec:genover} the overamped equation is instead obtained directly from the Generalized one, by simultaneously removing the momentum and the heat bath variables. Next, in Sec. \ref{sec:overbath}, we derive an overdamped equation for a system coupled to a set of heat bath variables. The heat baths are finally erased from the latter description, in Sec. \ref{sec:over2}, so that the standard overdamped equation is obtained through a suitable time scaling. Conclusions are drawn in Sec. \ref{sec:concl}.

\section{The Generalized Langevin Equation}
\label{sec:GLE}

The Generalized Langevin Equation (GLE) is given by \cite{chandler1987introduction, Pavl}
\begin{subequations}
\label{eq: gGLE}
\begin{align}
  dq&=\frac{p}{m}\, dt\\
  dp&=-\nabla U(q)\, dt-\nu\frac{p}{m}\, dt-\frac{1}{m}\int_{0}^t K(t-s) p(s)\,ds\, dt+F(t)\, dt+\sqrt{2\nu\beta^{-1}}\, d W(t).
\end{align}    
\end{subequations}
In the above equation, $q$ and $p$ denote the particle's position and momentum; $m>0$ is the mass of the particle, $U$ is a confining external potential; $\nu\geq 0$ is the viscous drag coefficient, $K(t)$ is a memory kernel that models the delayed drag effects \mcadded{exerted} by the fluid on the potential; $F(t)$ is a mean zero, stationary Gaussian process with autocovariance function, which is related to the memory kernel via the Fluctuation-Dissipation theorem \cite{Kubo66,Zwanzig,Vulp,LucCol12,Pavl}, i.e. $\mathbb{E}[F(t)F(s)]=K(|t-s|)$; and finally $W(t)$ is a standard Brownian motion. The GLE \eqref{eq: gGLE}\AM{, endowed with suitable initial conditions,} describes the motion of a microparticle \AM{in} a thermally fluctuating viscoelastic fluid\AM{. It }  has been employed in various applications such as surface diffusion 
and polymer dynamics, see the aforementioned monographs \cite{chandler1987introduction, Pavl} for further information on the GLE, including its derivation from a mechanical model coupled with a thermal reservoir. 

In general, due to the presence of the memory kernel, Eq. \eqref{eq: gGLE} describes a non-Markovian process. However, when $K(t)$ has the form of a finite sum of exponentials, viz.
\begin{equation}
    K(t)=\sum_{i=1}^M \lambda_i^2 e^{-\alpha_i t}, \quad t\geq 0
\end{equation}
then it has been proved that the non-Markovian GLE can be equivalently formulated as a Markovian system by introducing $M$ auxiliary variables \cite{kupferman2004fractional,Zwanzig}. More precisely, the augmented Markovian system takes the form
\begin{subequations}\label{eq:GLE0}
\begin{align}
  d q &= \frac{p}{m}\, dt , \label{eq:GLE-1} \\
  d p &= - \nabla_{q}U(q) \, dt -\nu\frac{p}{m}\, dt + \sum^{M}_{k=1} \lambda_k z_{k} \, dt+\sqrt{2\nu\beta^{-1}}\, dW_0(t)\label{eq:GLE-2}\,\\
  d z_{k} &= - \lambda_k \frac{p}{m}\, dt - \alpha_k z_{k} \, dt + \sqrt{ 2 \alpha_k \beta^{-1} } d W_{k}(t) \, , \quad k=1,\dots,M, \label{eq:GLE0-3}
\end{align}
\end{subequations}
where \mcadded{$\{z_k\}_{k=1}^M$} are augmented variables modelling the heat 
bath \AM{and} \mcadded{$\{W_k\}_{k=0}^M$} are independent standard Wiener processes. The system \eqref{eq:GLE0}, which we still refer to as the Generalized Langevin dynamics, has been studied extensively in the literature by many authors, see \cite{schuss2009theory, ottobre2011asymptotic,nguyen2018small,leimkuhler2022efficient, duong2022accurate, duong2023asymptotic} and references therein. In particular,  \AM{for suitable scaling of the involved parameters}, it has been proved that the GLE \eqref{eq:GLE0} can be cast into the framework of fast-slow dynamics. \AM{Using such a framework}, one can derive the classical \AM{structure of both the underdamped  and overdamped Langevin dynamics} by passing to \AM{suitable} limits \AM{in} the scaling parameter; \AM{for details, we refer the reader to \cite{schuss2009theory, ottobre2011asymptotic,nguyen2018small,duong2023asymptotic}}. We formally show these asymptotic limits in the following sections. 

\subsection{Model reduction for Generalized Langevin dynamics}

\begin{figure}
     \centering
         \includegraphics[width=\textwidth]{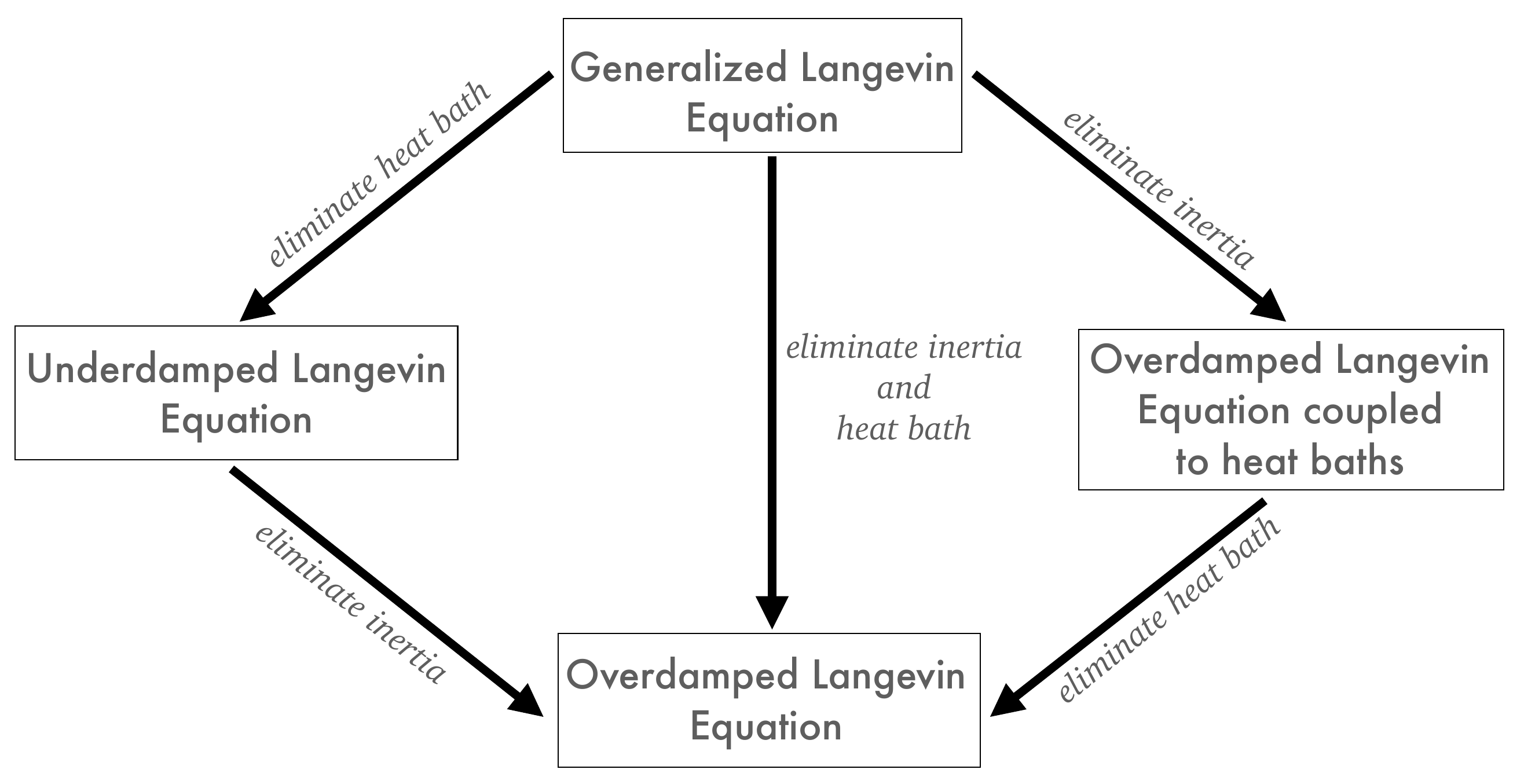}
        \caption{The stairs of reduction for the Generalized Langevin Equation given in Eqs. \eqref{eq:GLE0}. The reduction procedure detailed in the text guarantees that the diagram above is commutative, under a suitable time scaling.}
        \label{fig:diagram}
\end{figure}

In this paper, we carry out three different reduction paths for the GLE \eqref{eq:GLE0} equipped with quadratic confining potentials, in which we do not assume \mcadded{a priori} the presence of a \mcadded{time}-scale separation. The procedure is summarized in Figure \ref{fig:diagram}. The first path consists in \mcadded{deriving} the underdamped Langevin dynamics \mcadded{described by} the only position and momentum variables $\{q,p\}$ by eliminating the heat bath variables $\{z_k\}_{k=1}^M$ (see the top left arrow in Fig. \ref{fig:diagram}).
In the second path, we instead eliminate the inertia (i.e., the momentum variable $p$) to obtain a reduced model described by the position $q$ and heat baths variables $\{z_k\}_{k=1}^M$ (top right arrow in Fig. \ref{fig:diagram}). Using the same approach, we may eventually contract the description even further, by removing the inertia $p$ in the first reduced model (bottom left arrow in Fig. \ref{fig:diagram}) or the heat baths $\{z_k\}_{k=1}^M$ in the second one (bottom right arrow). We then show that the both reduction paths lead to the same reduced overdamped Langevin dynamics for the remaining position variable $q$. In the third path, we demonstrate that the latter reduced system can also be obtained from the GLE directly by removing both inertia and heat bath variables simultaneously (the central vertical downward arrow). In other words, the reduction diagram portrayed in Fig. \ref{fig:diagram} is commutative. 

Along each step of the reduction path, we employ the same \AM{tool -- the Invariant Manifold method. This} constitutes the backbone of a variety of reduction procedures exploited in classical kinetic theory \cite{GorKar05} and  \AM{with Fluctuation-Dissipation relations}, whose use in the set-up of model reduction was recently described in \cite{CM22,CDM22,CDM23}. Thereby, starting \AM{off} from a set of SDEs with additive noise and selecting a set of collective variables, a reduced model is derived in two steps: (i) the
deterministic component of for the reduced dynamics is obtained using the classical Invariant Manifold method, then (ii) the diffusion terms are determined by enforcing the Fluctuation-Dissipation Theorem. 
For the stochastic linear systems considered in this paper, the Fluctuation-Dissipation Theorem establishes a relation between the diffusion matrix, the transport matrix, and the stationary covariance matrix, expressed by a so-called Lyapunov equation \cite{Risken,Pavl}. We employ such relation twice, stipulating the scale invariance of the stationary covariance matrices. Thus, by construction, our approach paves the way to a multiscale connection between the transport and diffusion terms in \AM{both} the original and reduced dynamics.
In each of the considered reduction steps, we rescale the system through a parameter $\varepsilon$, which controls the time-scale separation between the ``fast'' and the ``slow'' variables. A perfect time-scale separation is achieved in the limit $\varepsilon\rightarrow 0$, where the standard Langevin equations encountered in the literature are recovered. There are two essential advantages coming from the considered rescaling procedure. The first one is that it provides a meaningful set-up to solve the Invariance Equations. The latter, which constitute a central tool in the derivation of hydrodynamics from kinetic theory \cite{Kar02,colan07,colan07b,colan08,colan09}, quantify the following conceptual recipe: the vector field of the original (deterministic) dynamics coincides with its projection onto a manifold parameterized by a finite set of slow variables, see also \cite{Gor04,GorKar}. Introducing an expansion of the fast variables in powers of $\varepsilon$ (that hence resembles the classical Chapman-Enskog expansion known in kinetic theory of gases \cite{CC}) makes it possible to solve iteratively the Invariance Equations.
The second advantage is that this procedure makes it possible to recover the aforementioned asymptotic limits as the leading order in the Chapman-Enskog expansion. Namely, our iterative solution of the Invariance Equations provides finite order corrections to the limiting dynamics, that address a perfect time-scale separation. We point out that our approach might also work without formally introducing the scaling parameter $\varepsilon$. Indeed, as discussed in \cite{GorKar}, one could set this parameter to unity (i.e., letting $\varepsilon=1$), and in some cases it is even possible to absorb it into some properly rescaled time and space variables, see e.g. \cite{colan07,colan07b,GorKar}.
In the following sections, for the simplicity and readability of the presentation, we  provide  detailed calculations for the case of one heat bath ($M=1$), but our approach is applicable to the general case of \mcadded{arbitrarily large, but finite,} number of heat baths; see Remark \ref{rem: arbitrary M} for \AM{further discussions}.

\section{From the GLE to the underdamped Langevin dynamics}
\label{sec:under}

In this section, we formally show, first, that the underdamped Langevin dynamics can be obtained from the GLE in the limit of \mcadded{perfect time-scale separation}, which has already been proved rigorously in \cite{ottobre2011asymptotic}. Later we perform our reduction procedure by eliminating the heat bath variables $\{z_k\}_{k=1}^M$ (top left arrow in Fig. \ref{fig:diagram}), thus obtaining higher order corrections to the underdamped Langevin dynamics.

\subsection{Formal $\varepsilon\rightarrow 0$ limit}

Rescaling the coefficients in \eqref{eq:GLE0} according to \AM{the rule } $\lambda_k\rightarrow \lambda_k/\varepsilon,~\alpha_k\rightarrow \alpha_k/\varepsilon^2$, we obtain the following rescaled generalized Langevin dynamics
\begin{subequations}\label{eq:rescaled-GLE}
\begin{align}
  d \qe &= \frac{\pe}{m}\, dt , \label{eq:rGLE-1} \\
  d \pe &= - \nabla_{q}U(\qe) \, dt-\nu \frac{\pe}{m}\, dt + \sum^{M}_{k=1} \frac{\lambda_k}{\varepsilon} \ze_{k} \, dt+\sqrt{2\nu\beta^{-1}}\, dW_0(t)\label{eq:rGLE-2} \\
  d \ze_{k} &= - \frac{\lambda_k}{\varepsilon}\frac{\pe}{m}\, dt - \frac{\alpha_k}{\varepsilon^2} \ze_{k} \, dt + \sqrt{ 2\frac{\alpha_k}{\varepsilon^2} \beta^{-1} } d W_{k}(t) \, , \quad k=1,\dots,M , \label{eq:rGLE-3}
\end{align}
\end{subequations}
where\mcadded{, as previously stated,} $\{W_k\}_{k=0}^M$ are independent standard Wiener processes.

\AM{Next,} we formally derive the limiting system as $\varepsilon\rightarrow 0$ from \eqref{eq:rescaled-GLE}; \AM{we refer the reader} to \cite{ottobre2011asymptotic} for a rigorous proof \AM{of this passage to the limit}. From \eqref{eq:rGLE-3}, we get
 \begin{align*}
  \frac{1}{\ve}\ze_k \, dt&=\frac{\ve}{\alpha_k}\Big[d\ze_k-\frac{\lambda_k}{\varepsilon}\frac{\pe}{m}\, dt + \sqrt{ 2\frac{\alpha_k}{\varepsilon^2} \beta^{-1} } d W_{k}(t)\Big] 
  \\&=\frac{\ve}{\alpha_k}d\ze_k-\frac{\lambda_k}{\alpha_k}\frac{\pe}{m}\,dt+\sqrt{\frac{2\beta^{-1}}{\alpha_k}}\, d W_{k}(t).
     \end{align*}
Substituting this expression into \eqref{eq:rGLE-2} yields
\begin{align*}
d \pe &= - \nabla_{q}U(\qe) \, dt + \ve\sum^{M}_{k=1}\frac{\lambda_k}{\alpha_k}d\ze_k-\Big(\nu+\sum_{k=1}^M\frac{\lambda_k^2}{\alpha_k}\Big)\frac{\pe}{m}\,dt+\Big(\sqrt{2\nu\beta^{-1}}\, dW_0(t)+\sum_{k=1}^M\sqrt{\frac{2\lambda_k^2\beta^{-1}}{\alpha_k}}\, d W_{k}(t)\Big) 
\\&=- \nabla_{q}U(\qe) \, dt + \ve\sum^{M}_{k=1}\frac{\lambda_k}{\alpha_k}d\ze_k-\gamma\frac{\pe}{m}\,dt+\sqrt{2\gamma\beta^{-1}}\, dW_t,
\end{align*}
where the friction coefficient $\gamma$ is defined by
\begin{equation}
  \gamma = \nu+\sum^{M}_{k=1} \frac{\lambda^{2}_{k}}{\alpha_{k}} \,.
  \label{gamma}
\end{equation}
\AM{To reach this point,} we have used the fact that
\[
\sqrt{\nu}\, dW_0(t)+\sum_{k=1}^M\sqrt{\frac{\lambda_k^2}{\alpha_k}}\, d W_{k}(t)
\]
is a Brownian motion with mean zero and variance $\gamma$. \AM{Hence, this} can be written as $\sqrt{\gamma}W(t)$, where $W(t)$ is \AM{the} standard Brownian motion. As $\varepsilon\rightarrow 0$, \AM{it yields that} the process $(q^\varepsilon, p^\varepsilon)$ converges to
\begin{subequations}
\label{eq:LE0}
\begin{align}
d q &= \frac{p}{m}\, dt , \label{eq:LE0-1} \\
d p &= - \nabla_{q}U(q) \, d t - \gamma \frac{p}{m}\, dt + \sqrt{2\gamma\beta^{-1}} d W(t) \, , \label{eq:LE0-2}
 \end{align}
  \end{subequations}
where the friction coefficient $\gamma$ is defined in \eqref{gamma}
and $W$, again, is a standard Wiener process. 
\subsection{Elimination of the heat bath variables}
We start \AM{off now} from the rescaled generalized Langevin dynamics \eqref{eq:rescaled-GLE}.
In the sequel, we denote by $\langle X \rangle:=\mathbb{E}[X]$ the expected value of the random variable $X$.
We look for a reduced description of the model described by Eq. \eqref{eq:GLE0} subject a harmonic potential $U(x)=1/2m\omega^2 q^2$, obtained by eliminating the set of heat bath variables $\{z_k\}_{k=1}^M$. 
We fix $M=1$ in Eq. \eqref{eq:GLE0}, then rescale $\lambda\rightarrow \lambda/\varepsilon$, $\alpha\rightarrow \alpha/\varepsilon^2$ and finally rescale also time as $t \rightarrow \varepsilon^2 t$. Noting that $dW(\varepsilon^2 t)=\varepsilon dW(t)$, we can rewrite the original dynamics as a linear system of SDEs:
\begin{equation}
d{\mathbf{z}}=\mathbf{M}_{\varepsilon} \mathbf{z} dt+\boldsymbol{{\sigma}}_{\varepsilon} d\mathbf{W}(t)   \, , \label{original2}
\end{equation}
where $\mathbf{z}=(q,p,z)^T$,  $d\mathbf{W}(t)=(0,dW_0(t),dW(t))^T$,
\begin{equation}
    \mathbf{M}_{\varepsilon}= \begin{pmatrix}
0 & \varepsilon^2/m  & 0  \\
-\varepsilon^2 m \omega^2 & -\varepsilon^2\nu/m  & \varepsilon \lambda  \\
0 & -\varepsilon \lambda/m  & -\alpha
\end{pmatrix} \, , \label{Mor}
\end{equation}
and 
\begin{equation}
    \boldsymbol{{\sigma}}_{\varepsilon}= \begin{pmatrix}
0 & 0  & 0  \\
0 & \varepsilon \sqrt{2 \nu/\beta}  & 0  \\
0 & 0  & \sqrt{2 \alpha/\beta}
\end{pmatrix} \, . \label{sigmaor}
\end{equation}
We target a reduced dynamics possessing the structure
\begin{equation}
    \label{redSDE}
d\mathbf{x}= \mathbf{M}_{r}(\varepsilon) \mathbf{x}\, dt + \boldsymbol{{\sigma}}_{r}(\varepsilon)\, d\mathbf{W}_r(t) \; ,
\end{equation}
with $\mathbf{x}=(q,p)^T$ and $d\mathbf{W}_r(t)=(dW_0(t),dW(t))^T$. The $2\times2$ matrices $\mathbf{M}_{r}$ and $\boldsymbol{{\sigma}}_{r}$, whose dependence on $\varepsilon$ is explicitly indicated in Eq. \eqref{redSDE}, will be determined exploiting the Invariant Manifold method and the Fluctuation-Dissipation Theorem (FDT).
We shall address the corresponding derivation in the coming sections.

\subsection{Invariance Equations and the Chapman-Enskog expansion}\label{sec:inveq}

Profiting from the linearity of the system \eqref{Mor}, we seek a closure \AM{relation} of the form:
\begin{eqnarray}
    \langle z \rangle &=& a(\varepsilon) \langle q \rangle + b(\varepsilon) \langle p \rangle \; , \label{clos2}   
\end{eqnarray}
where $a,b$ are regarded as functions of $\varepsilon$ (the other parameters of the model are considered as fixed).
One can \AM{then} compute the time derivative of $\langle z \rangle$ in two different ways; \AM{see \cite{GorKar05,CM22,CDM22} for similar arguments}. In the first, one uses directly the bottom row of \eqref{Mor}, relying also on the closure \eqref{clos2}. Alternatively, one considers the evolution of the fast variable $\langle z \rangle$ as driven by the slow variables $\langle q \rangle$ and $\langle p \rangle$. This way, the chain rule applies and, via \eqref{clos2}, this gives $\langle \dot{z}\rangle=a(\varepsilon) \, \langle\dot{q}\rangle+ b(\varepsilon) \, \langle \dot{p}\rangle$, where the time derivative of the slow variables is picked from the top and central rows of \eqref{Mor}. Within the framework of Invariant Manifold the two foregoing expressions of the time derivative of $\langle z \rangle$ coincide, which thus leads to two 
\textit{Invariance Equations} to be solved for the unknown functions $a(\varepsilon),b(\varepsilon)$:
\begin{subequations}
\label{IE}
\begin{eqnarray}
 -\varepsilon^2m \w^2 b+\varepsilon \lambda  ab + \alpha a  &=& 0 \label{IE1} \; , \\
\varepsilon^2 a-\varepsilon^2 \nu b +\varepsilon\lambda m b^2+ \varepsilon\lambda  +\alpha m b &=& 0  \; .\label{IE2}
\end{eqnarray}
\end{subequations}
The closure \AM{relation} \eqref{clos2} implies that
\begin{equation}
    \mathbf{M}_{r}(\varepsilon)=\varepsilon^2\begin{pmatrix}
0 & 1/m \\
-\Omega^2(\varepsilon) & -\Gamma(\varepsilon)
\end{pmatrix} \, , \label{Mred}
\end{equation}
with
\label{gener}
\begin{equation}
 \Omega^2(\varepsilon)= m\omega^2-\varepsilon^{-1}\lambda\, a(\varepsilon) \qquad , \qquad
 \Gamma(\varepsilon) = \frac{\nu}{m} -\varepsilon^{-1}\lambda\, b(\varepsilon)\, ,
\end{equation}
where $a(\varepsilon),b(\varepsilon)$ solve the system \eqref{IE1}--\eqref{IE2}.
Next, call \begin{equation}
    \xi^{\pm}(\varepsilon):=-\varepsilon^2\frac{\Gamma(\varepsilon)\pm\sqrt{\Gamma^2(\varepsilon)^2-4 \Omega^2(\varepsilon)/m}}{2}  \label{eigenval}
\end{equation}
the two eigenvalues of the matrix $\mathbf{M}_r$  in \eqref{Mred}.
Among the different sets of solutions $\{a(\varepsilon), b(\varepsilon)\}$ of Eqs. \eqref{IE}, the relevant ones are continuous functions fulfilling the asymptotic behavior:
\begin{equation}
  \lim_{\varepsilon\rightarrow 0}\xi^{\pm}(\varepsilon)=0 \; .\label{req}  
\end{equation}
The condition in \eqref{req} guarantees that, out of the full set of eigenvalues of the matrix $\mathbf{M}_{\varepsilon}$, the subset of eigenvalues retained in the reduced description are just the ``slow'' ones, namely those which vanish as $\varepsilon\rightarrow 0$.
To proceed further, we consider approximate solutions of the system \eqref{IE1}--\eqref{IE2} obtained through the Chapman-Enskog (CE) method.
One instructive way to perform the CE procedure amounts to expanding the coefficients $a(\varepsilon),b(\varepsilon)$ in powers of the parameter $\varepsilon$, viz.: 
\begin{equation}
\label{expan2}
a(\varepsilon)=\sum_{j=0}^{\infty}\varepsilon^j a_j  \qquad , \qquad b(\varepsilon)=\sum_{j=0}^{\infty}\varepsilon^j b_j \; ,
\end{equation} 
with $a_j, b_j$, $j=1,\dots,\infty$ denoting some real-valued coefficients,
and by then inserting the expansions \eqref{expan2} into the Invariance Equations \eqref{IE1}--\eqref{IE2}, which can thus be solved at any order of $\varepsilon$ \cite{GorKar05,GorKar}. 
A direct calculation yields in this case
\begin{equation}
   a(\varepsilon )=o(\varepsilon) \qquad , \qquad b(\varepsilon )=-\frac{\lambda}{m\alpha}\varepsilon+o(\varepsilon) \; .
      \label{1stb}
   \end{equation}

               

\subsection{The Fluctuation-Dissipation Theorem}
\label{sec:FDT}

Let us now turn to the computation of the matrix $\boldsymbol{{\sigma}}_{r}$ via the FDT.


We denote by $\mathbf{\Sigma}_{\varepsilon}=\frac{1}{2}\boldsymbol{{\sigma}}_{\varepsilon} \boldsymbol{{\sigma}}_{\varepsilon}^T$ the diffusion matrix and by $\mathbf{R}(t)$  the covariance matrix of the dynamics given in \eqref{original2}. 
One thus has
\begin{equation}
\mathbf{R}(t):=
\begin{pmatrix}
\hspace{60pt}\mathbf{R}_0(t)  & & \rvline &
  \begin{matrix}
  \langle  \delta q(t) \delta z(t) \rangle  \\
   \langle  \delta p(t) \delta z(t) \rangle  
  \end{matrix}
   \\
\hline
  \langle  \delta z(t) \delta q(t) \rangle  &  \langle  \delta z(t) \delta p(t) \rangle  & \rvline &  \langle  \delta z(t)^2 \rangle 
 
\end{pmatrix}
\label{Ror} \; ,
\end{equation}
where $\mathbf{R}_0(t)$ is the $2\times2$ matrix defined as
\begin{equation}
\mathbf{R}_0(t):=
\left(
    \begin{matrix}
\langle (q(t)-q(0))^2 \rangle & \langle (q(t)-q(0))(p(t)-p(0)) \rangle \\
\langle (p(t)-p(0))(q(t)-q(0)) \rangle & \langle(p(t)-p(0))^2\rangle 
\end{matrix}
\right)
\label{Rbl0} \; , 
\end{equation}
with $\delta q(t)=q(t)-\langle q(t)\rangle$.  
The stationary covariance matrix $\overline{\mathbf{R}}=\lim_{t\rightarrow\infty}\mathbf{R}(t)$ 
contains the block $\overline{\mathbf{R}_0}$, corresponding to the long time limit of $\mathbf{R}_0(t)$ in \eqref{Rbl0}.
The following relation represents an instance of the FDT \cite{Risken,Zwanzig,Pavl}, for it establishes a link between the matrices $\mathbf{\Sigma}_{\varepsilon}$, $\mathbf{M}_{\varepsilon}$ and $\overline{\mathbf{R}}$ in the form:
\begin{equation}
\mathbf{M}_{\varepsilon}\overline{\mathbf{R}}+\overline{\mathbf{R}} \mathbf{M}_{\varepsilon}^T=-2\mathbf{\Sigma}_{\varepsilon} \label{lyap} \; .
\end{equation}
Analogously, we define $\overline{\mathbf{R}}_r$ as the $2\times 2$ stationary covariance matrix of the reduced dynamics
\begin{equation}
\overline{\mathbf{R}}_r:= \lim_{t\rightarrow \infty}
\left(
    \begin{matrix}
\langle (q(t)-q(0))^2 \rangle & \langle (q(t)-q(0))(p(t)-p(0)) \rangle \\
\langle (p(t)-p(0))(q(t)-q(0)) \rangle & \langle(p(t)-p(0))^2\rangle
\end{matrix}
\right)
\label{covred} \; , 
\end{equation}
and $\mathbf{\Sigma}_r$ as the $2\times 2$ diffusion matrix of the reduced dynamics, see Eq. \eqref{redSDE}, such that $\mathbf{\Sigma}_r=\frac{1}{2}\boldsymbol{{\sigma}}_{r} \boldsymbol{{\sigma}}_{r}^T$.
It is possible to set-up the FDT also with the reduced dynamics, by letting:
\begin{equation}
\mathbf{M}_{r}\overline{\mathbf{R}_r}+\overline{\mathbf{R}_r} \mathbf{M}_{r}^T=-2\mathbf{\Sigma}_r \label{lyap2a} \; .
\end{equation}
In order to compute $\mathbf{\Sigma}_r$ through the knowledge of $\mathbf{M}_{\varepsilon}$ and $\mathbf{\Sigma}_{\varepsilon}$ we exploit an  algorithm comprising the following conceptual steps: (i) we solve Eq. \eqref{lyap} for $\overline{\mathbf{R}}$; (ii) we extract the block $\overline{\mathbf{R}_0}$ from the matrix $\overline{\mathbf{R}}$ and impose the scale-invariance of the correlation functions, i.e. we set $\overline{\mathbf{R}_0}=\overline{\mathbf{R}_r}$; (iii) we solve Eq. \eqref{lyap2a} for $\mathbf{\Sigma}_r$ (note that $\mathbf{M}_{r}$ is already known from the solution of the Invariance Equations \eqref{IE}). Application of the above algorithm yields
\begin{equation}
    \overline{\mathbf{R}_r}= \begin{pmatrix}
(m\beta \w^2)^{-1} & 0  \\
0 & m\beta^{-1}
\end{pmatrix} \, ,
\label{Rr}
\end{equation}
so that the reduced diffusion matrix attains the expression
\begin{equation}
     \mathbf{{\Sigma}}_{r}(\varepsilon)= \begin{pmatrix}
0 & -\varepsilon a(\varepsilon) \lambda/(2  \beta m \omega^2)  \\
-\varepsilon a(\varepsilon) \lambda/(2  \beta m \omega^2) & -\beta^{-1}\varepsilon[b(\varepsilon)m\lambda-\varepsilon\nu]
\end{pmatrix} \, .
\label{Sigred1}
\end{equation}
Upon inserting the CE expressions \eqref{1stb} into Eqs. \eqref{Mred} and \eqref{Sigred1}, then rescaling time back to $t \rightarrow \varepsilon^{-2} t$, and finally letting $\varepsilon\rightarrow 0$, one obtains 
\begin{equation}
    \mathbf{M}_{r}(0)=\begin{pmatrix}
0 & 1/m \\
-m\omega^2 & -\gamma/m
\end{pmatrix} \qquad
\text{and} \qquad
    \boldsymbol{{\sigma}}_{r}(0)= \begin{pmatrix}
0 & 0  \\
0 &\sqrt{2 \gamma/\beta}
\end{pmatrix} \, , \label{Msigred0}
\end{equation}
that properly match the structure of Eq. \eqref{eq:LE0} with $U(q)=1/2m\ \omega^2 q^2$ and $M=1$. 

               
\begin{remark}
\label{rem: arbitrary M}
\AM{We point out that}, due to the additive structure of the heat bath variables in \eqref{eq:GLE0}, our approach can be  extended to the case where there is an arbitrarily finite number of them, that is $M>1$.  We perform here the calculations for the model reduction of this section; similar calculations can be done for the model reduction steps in the subsequent sections. To this end, we use the closure 
\begin{eqnarray*}
    \langle z_k \rangle &=& a_k(\varepsilon) \langle q \rangle + b_k(\varepsilon) \langle p \rangle,
\end{eqnarray*}
where $a_k,b_k$ are yet unknown functions of the parameters $\{\omega,\lambda_k,\alpha_k\}$ for $k=1,\ldots, M$. Thus $a_k$ and $b_k$ satisfy the Invariance Equation \eqref{IE1}-\eqref{IE2} where $\lambda$ and $\alpha$ are replaced by $\lambda_k$ and $\alpha_k$. The generalized coefficients $\Omega^2$ and $\Gamma$ in \eqref{Mred} become
\begin{align*}
 \Omega^2(\varepsilon)&=m\omega^2-\sum_{k=1}^M\frac{\lambda_k}{\varepsilon } ~a_k(\varepsilon)\ , \\
 \Gamma(\varepsilon) &= \frac{\nu}{m} -\sum_{k=1}^M\frac{\lambda_k}{\varepsilon} ~b_k (\varepsilon) \,.
\end{align*}
The Chapman-Enskog expansion in \eqref{1stb} for each $k$, $k=1,\ldots, M$, yields
\begin{equation*}
   a_k(\varepsilon )=o(\varepsilon) \; , \; b_k(\varepsilon )=-\frac{\lambda_k }{m\alpha_k}\varepsilon+o(\varepsilon) \; .
\end{equation*}
This leads to
\[
\mathbf{M}_r(\varepsilon)=\varepsilon^2\begin{pmatrix}
    0&1/m\\
    -\Omega^2(\varepsilon)&-\Gamma(\varepsilon)
\end{pmatrix}=\varepsilon^2\begin{pmatrix}
    0&1/m\\
    -m\omega^2+\sum_{k=1}^M\frac{\lambda_k}{\varepsilon } ~a_k(\varepsilon)&-\frac{\nu}{m} +\sum_{k=1}^M\frac{\lambda_k}{\varepsilon} ~b_k (\varepsilon).
\end{pmatrix}
\]
By rescaling time as $t \rightarrow \varepsilon^2 t$, and then passing to the limit $\varepsilon\rightarrow 0$, \AM{we obtain} 
\begin{equation*}
\mathbf{M}_r(0)=\begin{pmatrix}
    0&1/m\\
    -m\omega^2&-\gamma/m
\end{pmatrix}.
\end{equation*}
We perform similar computations for the stochastic terms. \AM{Consequently, we recover} the underdamped Langevin equation \eqref{eq:LE0} for \AM{the case} $M>1$. 
\end{remark}

\section{From the underdamped to the overdamped Langevin dynamics}
\label{sec:over}

In this section, we first formally show that the overdamped Langevin dynamics can be obtained from the underdamped Langevin dynamics \eqref{eq:LE0} 
\mcadded{through a proper scaling}.
We shall, then, carry out the proposed model reduction procedure by eliminating the inertial variable (see the arrow shown in the bottom left corner of Fig. \ref{fig:diagram}). 
To this aim, we start letting $\gamma\mapsto \gamma/\varepsilon$ and $t\mapsto t/\varepsilon$ in Eqs. \eqref{eq:LE0}, so as to obtain the following rescaled underdamped Langevin equation:
\begin{subequations}
\label{eq:LEsc}
\begin{align}
d \qe &= \frac{\pe}{\ve m}\, dt , \label{eq:LE-1} \\
d \pe &= - \frac{1}{\ve}\nabla_{q}U(\qe) \, d t - \frac{\gamma}{\ve^2} \frac{\pe}{m}\, dt +\frac{1}{\varepsilon}\sqrt{\frac{2\gamma}{\beta}} d W(t) \, . \label{eq:LE-2}
 \end{align}
  \end{subequations}
  \subsection{Formal $\varepsilon\rightarrow 0$ limit}
We can formally pass to the limit $\varepsilon\rightarrow 0$ in the system above. From \eqref{eq:LE-1} and\eqref{eq:LE-2} we have
\[
d\qe=\frac{\pe}{\ve m}\,dt=-\frac{\ve}{\gamma} d\pe -\frac{1}{\gamma}\nabla_q U(\qe)\,dt+\sqrt{\frac{2}{\gamma \beta}}\, dW(t).
\]
Letting $\ve\rightarrow 0$, we obtain the overdamped Langevin dynamics
\begin{equation}
dq=-\frac{1}{\gamma}\nabla_q U(q)\,dt +\sqrt{\frac{2}{\gamma \beta}}\, dW(t) \; .
\label{overLE}
\end{equation}

\subsection{Elimination of inertial terms}
\label{sec:untoover}

We dwell in more detail, here, with the classical reduction step leading to the overdamped Langevin equation \eqref{overLE} starting from Eq. \eqref{eq:LEsc} with $U(q)=1/2 m \omega^2 q^2$, by eliminating the inertial terms, cf.  \cite{pavliotis2008multiscale,CM22}. We start from the rescaled equations \eqref{eq:LEsc}, and rescale also time as $t \rightarrow \varepsilon^2 t$. In the sequel, we shall retain the notation of the previous section, as long as no ambiguity arises.
The starting dynamics is described by an equation sharing the structure of Eq. \eqref{original2}, where we now set $\mathbf{z}=(q,p)^T$,  $d\mathbf{W}(t)=(0,dW(t))^T$, and with
\begin{equation}
\mathbf{M}_{\varepsilon}=\begin{pmatrix}
0 & \varepsilon/m \\
-\ve m\omega^2 & -\gamma/m
\end{pmatrix} \qquad ,  
\qquad
    \boldsymbol{{\sigma}}_{\varepsilon}= \begin{pmatrix}
 0  & 0  \\
0 & \sqrt{2  \gamma\beta^{-1}}  \\
\end{pmatrix} \, . \label{or2}
\end{equation}
We target an overdamped equation written in the form
\begin{equation}
    dq = M_r(\varepsilon) dt + \sqrt{2 \Sigma_r(\varepsilon)}\ dW(t) \; ,
    \label{overin}
\end{equation}
with yet unknown functions $M_r(\varepsilon),\Sigma_r(\varepsilon)$.
We stipulate a linear closure \AM{relation} $\langle p \rangle = c(\varepsilon) \langle q \rangle$, which gives $M_r(\varepsilon)=\varepsilon c(\varepsilon)/m$.
The Invariance Equation hence reads
\begin{equation}
    \varepsilon c^2 + \gamma c+ \ve m^2\omega^2=0 \; ,
    \label{inveq2}
\end{equation}
and its only relevant solution corresponds to the root of the quadratic equation \eqref{inveq2} which vanishes for $\varepsilon\rightarrow 0$. The FDT leads to the expression $\Sigma_r(\varepsilon)=\varepsilon^3 c(\varepsilon)/(\beta m^2 \omega^2)$, whereas
the application of the CE expansion yields:
\begin{equation}
\label{expan3}
c(\varepsilon)=-\ve m^2 \omega^2 \gamma^{-1}+O(\varepsilon^2) \; .
\end{equation}
Inserting \eqref{expan3} in the expressions of $M_r(\varepsilon)$ and $\Sigma_r(\varepsilon)$, then rescaling time back as $t \rightarrow \varepsilon^{-2} t$ and finally letting $\varepsilon\rightarrow 0$, one obtains $M_r(0)=-m \omega^2/\gamma$ and $\Sigma_r(0)=2(\gamma \beta)^{-1}$,
thus reproducing the structure of Eq. \eqref{overLE}.

\section{From the generalized to the overdamped Langevin dynamics}
\label{sec:genover}

In this section, we formally show, first, that the overdamped Langevin dynamics can be obtained \textit{directly} from the GLE via an appropriate rescaling of the parameters. Then, we perform our reduction procedure by erasing the momentum $p$ and the heat bath variables $\{z_k\}_{k=1}^M$ simultaneously (see the central vertical downward arrow in Fig. \ref{fig:diagram}).
\subsection{Formal $\varepsilon\rightarrow 0$ limit}
Consider the following rescaled generalized Langevin dynamics, viz.:
\begin{subequations}\label{eq:rr-GLE}
\begin{align}
  d \qe &= \frac{\pe}{m \ve^2 }\, dt , \label{eq:rrGLE-1} \\
  d \pe &= - \frac{1}{\ve^2}\nabla_{q}U(\qe) \, dt-\nu \frac{\pe}{m \ve^4}\, dt + \sum^{M}_{k=1} \frac{\lambda_k}{\ve^3} \ze_{k} \, dt+\frac{1}{\ve^2}\sqrt{2\nu \beta^{-1}}\, dW_0(t), \label{eq:rrGLE-2} \\
  d \ze_{k} &= - \frac{\lambda_k}{m\ve^3}\pe\, dt - \frac{\alpha_k}{\ve^2} \ze_{k} \, dt + \sqrt{ 2\frac{\alpha_k}{\ve^2} \beta^{-1} } d W_{k}(t) \, , \quad k=1,\dots,M . \label{eq:rrGLE-3}
\end{align}
\end{subequations}
We now formally derive the limiting system as $\ve\rightarrow 0$. We have
\[
\frac{\ze_k}{\ve^3}\,dt\overset{\eqref{eq:rrGLE-3}}{=}\Big[\frac{1}{\alpha_k \ve}d\ze_k-\frac{\lambda_k}{m\alpha_k \ve^4}\pe\,dt+\sqrt{\frac{2\beta^{-1}}{\alpha_k}}\frac{1}{\ve^2}dW_k\Big].
\]
Substituting this into \eqref{eq:rrGLE-2}
\begin{align*}
 d\pe&=   - \frac{1}{\ve^2}\nabla_{q}U(\qe) \, dt + \sum^{M}_{k=1}\frac{\lambda_k}{\alpha_k \ve}d\ze_k-\Big(\nu+\sum_{k=1}^M\frac{\lambda_k^2}{\alpha_k}\Big)\frac{\pe}{m \ve^4}\,dt+\sqrt{2\nu\beta^{-1}}\frac{1}{\ve^2}\, dW_0(t)+\sum_{k=1}^M \sqrt{\frac{2\beta^{-1}\lambda_k^2}{\alpha_k}}\frac{1}{\ve^2}dW_k
 \\&=- \frac{1}{\ve^2}\nabla_{q}U(\qe) \, dt + \sum^{M}_{k=1}\frac{\lambda_k}{\alpha_k \ve}d\ze_k-\gamma\frac{\pe}{ m\ve^4}\,dt+\frac{1}{\ve^2}\sqrt{2\gamma \beta^{-1}}dW(t),
\end{align*}
where $\gamma$ is defined in \eqref{gamma} and $W(t)$ is a standard Wiener process.
It follows that
\[
\frac{\pe}{m\ve^2}\, dt=\frac{\ve^2}{\gamma}d\pe-\frac{1}{\gamma}\nabla_q U(\qe)\,dt+\frac{\ve}{\gamma}\sum_{k=1}^M\frac{\lambda_k}{\alpha_k}d\ze +\sqrt{\frac{2\beta^{-1}}{\gamma}}\,dW(t).
\]
Substituting into \eqref{eq:rrGLE-1} yields
\begin{align*}
 d \qe=\frac{\ve^2}{\gamma}d\pe-\frac{1}{\gamma}\nabla_q U(\qe)\,dt+\frac{\ve}{\gamma}\sum_{k=1}^M\frac{\lambda_k}{\alpha_k}d\ze +\sqrt{\frac{2\beta^{-1}}{\gamma}}\,dW(t)
\end{align*}
Thus as $\varepsilon\rightarrow 0$, the process $\qe$ converges to 
\begin{equation}
dq=-\frac{1}{\gamma}\nabla_q U(q)\,dt+\sqrt{\frac{2\beta^{-1}}{\gamma}}dW(t) \;,
\label{overd}
\end{equation}
which is precisely the overdamped Langevin dynamics \eqref{overLE}.
\subsection{Elimination of inertial terms and heat bath variables}
\label{sec:gentoover}


In Eqs. \eqref{eq:rr-GLE} we fix again $M=1$ and $U(q)=1/2 m \omega^2 q^2$. Upon rescaling time as $t\rightarrow \varepsilon^4 t$, we can rewrite the original dynamics in the form expressed by Eq. \eqref{original2}, where we set $\mathbf{z}=(q,p,z)^T$,  $d\mathbf{W}(t)=(0,dW_0(t),dW(t))^T$,
\begin{equation}
    \mathbf{M}_{\varepsilon}= \begin{pmatrix}
0 & \varepsilon^2/m  & 0  \\
-\varepsilon^2 m \omega^2 & -\nu/m  & \varepsilon \lambda  \\
0 & -\varepsilon \lambda/m  & -\varepsilon^2 \alpha
\end{pmatrix} \, , \label{Mor2}
\end{equation}
and 
\begin{equation}
    \boldsymbol{{\sigma}}_{\varepsilon}= \begin{pmatrix}
0 & 0  & 0  \\
0 &  \sqrt{2  \nu/\beta}  & 0  \\
0 & 0  &  \varepsilon \sqrt{2  \alpha/\beta}
\end{pmatrix} \, . \label{sigmaor2}
\end{equation}
Note that in \eqref{sigmaor2} we have exploited the scaling $dW(\varepsilon^4 t)=\varepsilon^{2}dW(t)$.
We seek an overdamped dynamics in the form expressed by Eq. \eqref{overin}. We thus introduce the closure relations:
\begin{equation*}
    \langle p \rangle = a(\varepsilon) \langle q \rangle \; , \;  \langle z \rangle = b(\varepsilon) \langle q \rangle  \; , 
\end{equation*}
which lead to the expression $M_r(\varepsilon)=\varepsilon^2 a(\varepsilon)/m$ and, via the FDT, also to $\Sigma_r(\varepsilon)=-\varepsilon^2 a(\varepsilon)/(\beta m^2 \w^2)$.
The functions $a(\varepsilon),b(\varepsilon)$ are obtained as solutions of the Invariance Equations:
\begin{subequations}
\label{inveq3}
\begin{eqnarray}
 \varepsilon^2 m^2 \w^2+\nu a -\varepsilon \lambda m b +\varepsilon^2 a^2 &=&0 \label{IE3a}\; , \\
\lambda a +\varepsilon m \alpha b+ \varepsilon a b &=& 0 \label{IE3b} \; .
\end{eqnarray}
\end{subequations}
Among the different sets of solutions $\{a(\varepsilon),b(\varepsilon)\}$ of Eqs. \eqref{inveq3}, the relevant ones are those which let $M_r$ vanish as $\varepsilon\rightarrow 0$. 
The application of the CE expansion yields:
\begin{equation}
\label{expan4}
a(\varepsilon)=-\varepsilon^2 m^2 \omega^2 \gamma^{-1}+o(\varepsilon^2) \qquad , \qquad b(\varepsilon)=\varepsilon m \omega^2 \lambda \alpha\gamma^{-1}+o(\varepsilon) \; .
\end{equation}
Inserting the expansions \eqref{expan4} in the expressions of $M_r(\varepsilon)$ and $\Sigma_r(\varepsilon)$, then rescaling time back as $t \rightarrow \varepsilon^4 t$ and finally letting $\varepsilon\rightarrow 0$, one gets $M_r(0)=-m \omega^2/\gamma$ and $\Sigma_r(0)=(\gamma \beta)^{-1}$,
hence recovering the structure of Eq. \eqref{overLE}, and also matching the overdamped equation previously obtained in Sec. \ref{sec:untoover}.

\section{From the generalized Langevin to the overdamped Langevin dynamics coupled to heat baths}
\label{sec:overbath}

In this section we consider the reduction path portrayed in the upper right corner of Fig. \ref{fig:diagram}, \mcadded{namely we eliminate the momentum variable from the GLE}. As in the previous sections, we first show that, under an appropriate scaling, one can derive a limiting dynamics that couples the position variable to the heat bath variables. \mcadded{Next, we shall carry out in detail our reduction procedure.}

\subsection{Formal $\varepsilon\rightarrow 0$ limit}
We consider the following rescaled GLE equation
\begin{subequations}\label{eq:r-GLE4}
\begin{align}
  d \qe &= \frac{\pe}{\ve m}\, dt , \label{eq:rGLE4-1} \\
  d \pe &= -\frac{1}{\ve} \nabla_{q}U(\qe) \, dt -\nu\frac{\pe}{m\ve^2}\, dt + \frac{1}{\ve}\sum^{M}_{k=1} \lambda_k \ze_{k} \, dt+\frac{\sqrt{2\beta^{-1}\nu}}{\ve}\, dW_0(t),\label{eq:rGLE4-2}\,\\
  d \ze_{k} &= - \lambda_k \frac{\pe}{\ve m}\, dt - \alpha_k \ze_{k} \, dt + \sqrt{ 2 \alpha_k \beta^{-1} } d W_{k}(t) \, , \quad k=1,\dots,M . \label{eq:rGLE4-3}
\end{align}
\end{subequations}
We define
\[
\ye_k:=\lambda_k\qe+ \ze_k, \quad k=1,\ldots, M.
\]
Then, it follows from \eqref{eq:rGLE4-1} and \eqref{eq:rGLE4-3} that
\begin{align*}
d\ye_k&= - \alpha_k \ze_{k} \, dt + \sqrt{ 2 \alpha_k \beta^{-1} } d W_{k}(t) 
\\&=\alpha_k\lambda_k \qe \, dt-\alpha_k \ye_k\, dt + \sqrt{ 2 \alpha_k \beta^{-1} } d W_{k}(t). 
\end{align*}
The GLE system \eqref{eq:r-GLE4} is equivalent to the following system in terms of $(\qe, \pe, \ye)$
\begin{subequations}
\label{eq:rGLE41}
\begin{align}
  d \qe &= \frac{\pe}{\ve m}\, dt , \label{eq:rGLE41-1} \\
  d \pe &= -\frac{1}{\ve} \nabla_{q}U(\qe) \, dt -\nu\frac{\pe}{m\ve^2}\, dt + \frac{1}{\ve}\sum^{M}_{k=1} \lambda_k (\ye_k-\lambda_k\qe) \, dt+\frac{\sqrt{2\beta^{-1}\nu}}{\ve}\, dW_0(t),\label{eq:rGLE41-2}\,\\
  d\ye_k &=\alpha_k\lambda_k \qe \, dt-\alpha_k \ye_k\, dt + \sqrt{ 2 \alpha_k \beta^{-1} } d W_{k}(t)\label{eq: rGLE41-3}.
\end{align}
\end{subequations}
We perform the reduction of this system by eliminating the inertia variable $\pe$. We formally derive the limiting system by letting $\ve\rightarrow 0$. From \eqref{eq:rGLE-1} and \eqref{eq:rGLE-2} we have
\begin{align*}
  d \qe = \frac{\pe}{\ve m}\, dt&=\frac{\ve}{\nu}\Big(-d\pe-\frac{1}{\ve} \nabla_{q}U(\qe) \, dt + \frac{1}{\ve}\sum^{M}_{k=1} \lambda_k (\ye_k-\lambda_k\qe) \, dt+\frac{\sqrt{2\beta^{-1}\nu}}{\ve}\, dW_0(t)\Big)
  \\&=-\frac{\ve}{\nu}d\pe-\frac{\nabla_q U(\qe)}{\nu}\, dt+\frac{1}{\nu}\sum_{k=1}^M \lambda_k (\ye_k-\lambda_k\qe)\,dt+\sqrt{\frac{2\beta^{-1}}{\nu}}\, dW_0(t).
\end{align*}
By sending $\ve\rightarrow 0$, we obtain a coupled system \AM{involving} $q$ and $y_k$
\begin{subequations}\label{eq:lGLE3}
\begin{align}
  d q &= -\frac{1}{\nu}\Big(\nabla_q U(q)+\sum_{k=1}^M\lambda_k^2 q\Big)\, dt+\frac{1}{\nu}\sum_{k=1}^M \lambda_k y_k\,dt +\sqrt{\frac{2\beta^{-1}}{\nu}}\, dW_0(t).\label{eq:lGLE3-1}\\
dy_k&=\alpha_k\lambda_k q \, dt - \alpha_k y_{k}\,dt + \sqrt{ 2 \alpha_k \beta^{-1} } d W_{k}(t) \, , \quad k=1,\dots,M \; . \label{eq:lGLE3-2}
\end{align}
\end{subequations}
Let
\[
\bar{U}(q):=U(q)+\frac{1}{2}\sum_{k=1}^M \lambda_k^2 q^2
\]
denote the effective potential. Then the limiting system can be written as
\begin{subequations}\label{eq:lGLE3f}
\begin{align}
  d q & = -\frac{1}{\nu}\nabla_q \bar{U}(q)\, dt+\frac{1}{\nu}\sum_{k=1}^M \lambda_k y_k\,dt+\sqrt{\frac{2\beta^{-1}}{\nu}}\, dW_0(t),
 \label{eq:lGLE3f-1}\\
dy_k&=\alpha_k\lambda_k q \, dt - \alpha_k y_{k}\,dt + \sqrt{ 2 \alpha_k \beta^{-1} } d W_{k}(t) \, , \quad k=1,\dots,M . \label{eq:lGLE3f-2}
\end{align}
\end{subequations}
This system directly couples the position variable $q$, \mcadded{characterizing} an overdamped Langevin dynamics, to the heat bath \mcadded{variables}. This kind of system has been studied by many authors in the literature in the context of dynamical systems driven by \mcadded{colored noise}, see e.g. \cite{haunggi1994colored}. Rigorous analysis for the above asymptotic limit has also been done in the literature recently \cite{duong2023asymptotic}.

In particular, when $U(q)=1/2 m\omega^2 q^2$, the above system boils down to
\begin{subequations}\label{eq:lGLE31}
\begin{align}
  d q &= -\frac{1}{\nu}\Big(m\omega^2+\sum_{k=1}^M\lambda_k^2\Big)q\, dt+\frac{1}{\nu}\sum_{k=1}^M \lambda_k y_k\,dt+\sqrt{\frac{2\beta^{-1}}{\nu}}\, dW_0(t), \\
dy_k&=\alpha_k\lambda_k q \, dt - \alpha_k y_{k}\,dt + \sqrt{ 2 \alpha_k \beta^{-1} } d W_{k}(t) \, , \quad k=1,\dots,M . \label{eq:GLE-3}
\end{align}
\end{subequations}
Let
\[
\zeta_0:=\frac{1}{\nu}\Big(m\omega^2+\sum_{k=1}^M\lambda_k^2\Big),\quad \eta_k:=\frac{1}{\nu}\lambda_k, \quad \zeta_k:=\alpha_k \lambda_k.
\]
Then, Eq. \eqref{eq:lGLE31} reads
\begin{subequations}
\label{eq:lGLE4}
\begin{align}
  d q &= - \zeta_0 q \, dt + \sum^{M}_{k=1} \eta_k y_{k} \, dt+\sqrt{\frac{2\beta^{-1}}{\nu}}\, dW_0(t) \label{eq:lGLE4-1} \\
  d y_{k} &= \zeta_k q \,dt- \alpha_k y_{k} \, dt + \sqrt{ 2 \alpha_k \beta^{-1} } d W_{k}(t) \, , \quad k=1,\dots,M . \label{eq:lGLE4-2}
\end{align}
\end{subequations}
\subsection{Elimination of inertial terms}
We start our reduction procedure from the rescaled GLE equation \eqref{eq:lGLE4}, where we rescale time as $t\rightarrow \varepsilon^2 t$.
The resulting dynamics can thus be cast in the structure of Eq. \eqref{original2}, where we now set $\mathbf{z}=(q,p,y)^T$,  $d\mathbf{W}(t)=(0,dW_0(t),dW(t))^T$,
\begin{equation}
    \mathbf{M}_{\varepsilon}= \begin{pmatrix}
0 & \varepsilon/m  & 0  \\
-\varepsilon (m\omega^2-\lambda^2) & -\nu/m  & \varepsilon \lambda  \\
\varepsilon^2 \lambda \alpha & 0  & -\varepsilon^2\alpha
\end{pmatrix}  \, , \label{Mor3}
\end{equation}
and 
\begin{equation}
    \boldsymbol{{\sigma}}_{\varepsilon}= \begin{pmatrix}
0 & 0  & 0  \\
0 & \sqrt{2 \nu/\beta}  & 0  \\
0 & 0  & \varepsilon \sqrt{2 \alpha/\beta}
\end{pmatrix} \, . \label{sigmaor3}
\end{equation}
We seek a reduced description described by Eq. \eqref{redSDE}, with $\mathbf{x}=(q,y)^T$ and $d\mathbf{W}_r(t)=(dW_0(t),dW(t))^T$. The considered closure takes here the form 
\begin{equation*}\langle p \rangle = a(\varepsilon) \langle q \rangle + b(\varepsilon) \langle y \rangle \; ,
\end{equation*}
which thus yields
\begin{equation}
    \mathbf{M}_{r}(\varepsilon)=\begin{pmatrix}
\varepsilon a(\varepsilon)/m & \varepsilon b(\varepsilon)/m \\
\varepsilon^2\lambda \alpha & -\varepsilon^2\alpha
\end{pmatrix} \, , \label{Mredc}
\end{equation}
whereas application of the FDT leads to
\begin{equation}
    \mathbf{\Sigma}_{r}(\varepsilon)=\begin{pmatrix}
-\frac{(a(\varepsilon) + \lambda b(\varepsilon)) \varepsilon}{m^2 \beta \omega^2} & -\frac{[(a(\varepsilon) + \lambda b(\varepsilon))\lambda+ b(\varepsilon) m \omega^2] \varepsilon}{2m^2 \beta \omega^2}  \\
-\frac{[(a(\varepsilon) + \lambda b(\varepsilon))\lambda+ b(\varepsilon) m \omega^2] \varepsilon}{2m^2 \beta \omega^2}  & \frac{\varepsilon^2\alpha}{\beta} 
\end{pmatrix} \, . \label{Sigredc}
\end{equation}
The Invariance Equations read:
\begin{subequations}
 \label{IE3}
\begin{eqnarray*}
 \varepsilon m^2 \w^2+\nu a+\varepsilon m \lambda^2+\varepsilon a^2 + \varepsilon^2 m\alpha \lambda b  &=& 0 \; , \\
\nu b-\varepsilon m\lambda+\varepsilon a b-\varepsilon^2 m \alpha b &=& 0 \; ,
\end{eqnarray*}   
\end{subequations}
while CE method returns the expressions
\begin{equation}
   a(\varepsilon )=-m \zeta_0 \varepsilon  + o(\varepsilon) \; , \; b(\varepsilon )=m\eta \varepsilon +o(\varepsilon) \; .
      \label{expan5}
   \end{equation}
Inserting the expansions \eqref{expan5} into \eqref{Mredc} and \eqref{Sigredc}, then rescaling time back as $t \rightarrow \varepsilon^{-2} t$ and finally letting $\varepsilon\rightarrow 0$ yields 
\begin{equation}
    \mathbf{M}_{r}(0)=\begin{pmatrix}
- \zeta_0 &  \eta \\
\zeta & -\alpha
\end{pmatrix} \, , \label{Mredc2}
\end{equation}
and
\begin{equation}
     \boldsymbol{{\sigma}}_{r}(0)=\begin{pmatrix}
\sqrt{2/(\beta \nu)} & 0  \\
0  & \sqrt{2 \alpha/\beta}
\end{pmatrix} \, . \label{sigredc2}
\end{equation}
The expressions \eqref{Mredc2}--\eqref{sigredc2} correctly recover the structure of Eqs. \eqref{eq:lGLE4}.

\section{From the overdamped Langevin-heat bath equation to the overdamped Langevin}
\label{sec:over2}

In this section, we carry out the last reduction step, shown in the \mcadded{bottom right corner} of Fig. \ref{fig:diagram}, \mcadded{namely we eliminate} the heat baths in the coupled overdamped Langevin-heat bath equation derived in the previous section. 
\subsection{Formal $\varepsilon\rightarrow 0$ limit}
Consider the following rescaled system of \eqref{eq:lGLE3}
\begin{subequations}
\label{eq:rlGLE4}
\begin{align}
  d \qe &= -\frac{1}{\nu}\nabla_q U(\qe)\, dt-\frac{1}{\nu}\sum_{k=1}^M \lambda_k^2\qe\, dt + \frac{1}{\nu}\sum^{M}_{k=1} \lambda_k\ye_{k} \, dt+\sqrt{\frac{2\beta^{-1}}{\nu}}\, dW_0(t),\label{eq:rlGLE4-1} \\
  d \ye_{k} &= \frac{\lambda_k \alpha_k}{\ve} \qe \,dt- \frac{\alpha_k}{\ve}\ye_{k} \, dt + \sqrt{2 \alpha_k \beta^{-1} \ve^{-1}} d W_{k}(t) \, , \quad k=1,\dots,M . \label{eq:rlGLE4-2}
\end{align}
\end{subequations}
From this system, we will derive the overdamped Langevin dynamics by letting $\varepsilon\rightarrow 0$. From \eqref{eq:rlGLE4-2}, we get
\[
\ye_k\,dt=-\frac{\ve}{\alpha_k} d\ye_k+\lambda_k\qe\,dt+\sqrt{\frac{2\beta^{-1}\ve}{\alpha_k}}\, dW_k(t).
\]
Substituting this back into \eqref{eq:rlGLE4-1} yields
\begin{align*}
d \qe& = - \frac{1}{\nu}\Big(\nabla_q U(\qe)+\sum_{k=1}^M\lambda_k^2\qe) \, dt + \frac{1}{\nu}\sum^{M}_{k=1} \lambda_k\Big(-\frac{\ve}{\alpha_k} d\ye_k+\lambda_k \qe\,dt+\sqrt{\frac{2\beta^{-1}\ve}{\alpha_k}}\, dW_k(t)\Big)+\sqrt{\frac{2\beta^{-1}}{\nu}}\, dW_0(t)
\\&=-\frac{1}{\nu} \nabla_q U(\qe)\,dt-\frac{\ve}{\nu}\sum^{M}_{k=1} \frac{\lambda_k }{\alpha_k} d\ye_k+\frac{1}{\nu}\sqrt{2\beta^{-1}\ve\sum_{k=1}^M\frac{\lambda_k^2}{\alpha_k}}\, dW_k(t)+\frac{1}{\nu}\sqrt{2\beta^{-1}\nu}\, dW_0(t).
\end{align*}
Next, upon sending $\ve\rightarrow 0$ we get
\begin{equation}
dq=-\frac{1}{\nu} \nabla_q U(q)\,dt+\sqrt{\frac{2\beta^{-1}}{\nu}}\, dW(t)\; ,\label{overdfin}
\end{equation}
which is exactly the overdamped Langevin dynamics after an appropriate rescaling of time, as it will be seen below.

\subsection{Elimination of heat bath variables}

\mcadded{Let us thus apply our reduction procedure to the Langevin-heat bath equation, in order to obtain an overdamped Langevin dynamics.}
To this aim, consider the rescaled system \eqref{eq:rlGLE4}.
We let $U(q)=1/2 m \omega^2 q^2$, rescale time as $t\rightarrow \varepsilon t$ and thus write the resulting dynamics in the form given by Eq. \eqref{original2}, in which we set $\mathbf{z}=(q,y)^T$, $d\mathbf{W}(t)=(dW_0(t),dW(t))^T$, and
\begin{equation}
    \mathbf{M}_{\varepsilon}= \begin{pmatrix}
-\varepsilon m\omega^2/\nu-\varepsilon\lambda^2/\nu & \varepsilon\lambda/\nu  \\
\lambda \alpha & -\alpha
\end{pmatrix} \,  \label{Mor2c}
\end{equation}
and
\begin{equation}
   \boldsymbol{{\sigma}}_{\varepsilon}= \begin{pmatrix}
\sqrt{\frac{2 \varepsilon}{\nu \beta}}  & 0  \\
0  & \sqrt{\frac{2 \alpha}{\beta}}
\end{pmatrix} \, .  \label{sigmac}
\end{equation}
Our target is an overdamped equation sharing the structure of Eq. \eqref{overin}.
We here exploit the closure relation $\langle y \rangle = c(\varepsilon) \langle q \rangle$, which leads to \begin{equation}
M_r(\varepsilon)=-\frac{\varepsilon (m\omega^2+\lambda^2 -c(\varepsilon)\lambda)}{\nu} \; ,
\label{Mrfin}
\end{equation}
and, via the FDT, also to
\begin{equation}
\Sigma_{r}(\varepsilon)=\frac{\varepsilon (-c(\varepsilon) \lambda+\lambda^2 +m\omega^2)}{m \beta \nu \omega^2} \; .
\label{Sigridfin}
\end{equation}
In this case the Invariance Equation reads:
\begin{equation}
\label{IE5}
\lambda \alpha \nu-\alpha c \nu+c\varepsilon m \omega^2+c \varepsilon\lambda^2 -\varepsilon c^2 \lambda   = 0 \; ,
\end{equation}
whereas the CE expansion yields $c(\varepsilon)=\lambda+O(\varepsilon)$.
Inserting the latter expression into \eqref{Mrfin} and \eqref{Sigridfin}, then rescaling time back as $t \rightarrow \varepsilon^{-1} t$ and by finally letting $\varepsilon \rightarrow 0$, one finds $M_r(0)=-m\w^2/\nu$ and $\Sigma_r(0)=1/(\beta \nu)$, which recover
the coefficients appearing in Eq. \eqref{overdfin}. 
Finally, a further rescaling of time $t \rightarrow \nu/\gamma\ t$, such that $dW(\nu/\gamma\ t)=\sqrt{\nu/\gamma}\, dW(t)$, guarantees that the overdamped equations obtained in Secs. \ref{sec:untoover} and \ref{sec:gentoover} are also properly matched. This hence proves the commutativity of the various reduction paths shown in the diagram of Fig. \ref{fig:diagram}.

\section{Conclusions}
\label{sec:concl}

We discussed the derivation of reduced Langevin dynamics, either underdamped or overdamped, with or without the coupling with heat bath variables, from a Generalized Langevin Equation, that stands as a paradigmatic model describing non-Markovian processes. \AM{In the approach presented here}, we rescaled the various Langevin dynamics through a parameter $\varepsilon$, controlling the time-scale separation between slow and fast variables. Our procedure is based on the method of the Invariant Manifold, which unveils finite order corrections to the standard Langevin equations known in the literature. Remarkably, these are properly recovered, within our formalism, in the regime of perfect time-scale separation, attained in the limit $\varepsilon\rightarrow 0$. The considered reduction procedure is also underpinned by the Fluctuation-Dissipation Theorem, which is exploited at different levels of description through the assumed scale-invariance of the stationary covariance matrices. Our method thus leads to an explicit form of the diffusion terms in the reduced dynamics, which match \AM{well-known formulas reported} in the literature when a perfect time-scale separation holds. Another noteworthy feature of our technique is that it proves to be independent of the chosen reduction path. That is, the same overdamped description can be obtained from the Generalized Langevin Equation either by erasing first the inertial terms and then the heat bath variables, or vice versa. This way, the diagram portrayed in Fig. \ref{fig:diagram} turns out being commutative.

Many relevant open questions still remain ahead. One is related, for instance, to the derivation of reduced descriptions, obtained from the Generalized Langevin Equations, through the \AM{perspective} of response theory, following the guidelines traced e.g. in \cite{Maes09,Maes10,Col11}. Another intriguing aspect concerns the analysis of Langevin dynamics coupled to heat baths of various species, that may thus evolve on different time scales. These topics will be \AM{a  matter} for future work.

\textbf{Acknowledgements}

This work was carried out under the auspices of the Italian National Group of Mathematical Physics.
MC thanks the PRIN 2022 project
``Mathematical modelling of heterogeneous systems"
(code 2022MKB7MM, CUP B53D23009360006). \AM{The research of MHD is} funded by the EPSRC Grant EP/V038516/1 and the a Royal International
Exchange Grant IES-R3-223047.

\bibliographystyle{alpha}
\bibliography{biblio}

\newcommand{\etalchar}[1]{$^{#1}$}
\begin{thebibliography}{BBMW10}

\bibitem[BAB{\etalchar{+}}23]{bianconi2023complex}
G.~Bianconi, A.~Arenas, J.~Biamonte, L.~D. Carr, B.~Kahng, J.~Kertesz,
  J.~Kurths, L.~L{\"u}, C.~Masoller, A.~Motter E, et~al.
\newblock Complex systems in the spotlight: next steps after the 2021 nobel
  prize in physics.
\newblock {\em Journal of Physics: Complexity}, 4(1):010201, 2023.

\bibitem[BBMW10]{Maes10}
M.~Baiesi, E.~Boksenbojm, C.~Maes, and B.~Wynants.
\newblock Nonequilibrium {Linear} {Response} for {Markov} {Dynamics}, {II}:
  {Inertial} {Dynamics}.
\newblock {\em Journal of Statistical Physics}, 139(3):492--505, 2010.

\bibitem[BMW09]{Maes09}
M.~Baiesi, C.~Maes, and B.~Wynants.
\newblock Nonequilibrium {Linear} {Response} for {Markov} {Dynamics}, {I}:
  {Jump} {Processes} and {Overdamped} {Diffusions}.
\newblock {\em Journal of Statistical Physics}, 137(5):1094--1116, 2009.

\bibitem[BOCW17]{benner2017model}
P.~Benner, M.~Ohlberger, A.~Cohen, and K.~Willcox.
\newblock {\em Model reduction and approximation: theory and algorithms}.
\newblock SIAM, 2017.

\bibitem[CC70]{CC}
S.\ Chapman and T.~G.\ Cowling.
\newblock {\em The {M}athematical {T}heory of {N}onuniform {G}ases}.
\newblock Cambridge University Press, New York, 1970.

\bibitem[CDM22]{CDM22}
M.~Colangeli, M.~H. Duong, and A.~Muntean.
\newblock A reduction scheme for coupled {Brownian} harmonic oscillators.
\newblock {\em Journal of Physics A: Mathematical and Theoretical}, 55:505002,
  2022.

\bibitem[CDM23]{CDM23}
M.~Colangeli, M.~H. Duong, and A.~Muntean.
\newblock Model reduction of brownian oscillators: quantification of errors and
  long-time behavior.
\newblock {\em Journal of Physics A: Mathematical and Theoretical},
  56(34):345003, 2023.

\bibitem[Cha87]{chandler1987introduction}
D.~Chandler.
\newblock {Introduction to Modern Statistical Mechanics}.
\newblock {\em Oxford University Press, Oxford, UK}, 5(449):11, 1987.

\bibitem[CKK07a]{colan07}
M.\ Colangeli, I.~V.\ Karlin, and M.\ Kr\"{o}ger.
\newblock From hyperbolic regularization to exact hydrodynamics for linearized
  {Grad}'s equations.
\newblock {\em Phys. Rev. E}, 75:051204, 2007.

\bibitem[CKK07b]{colan07b}
M.\ Colangeli, I.~V.\ Karlin, and M.\ Kr\"{o}ger.
\newblock Hyperbolicity of exact hydrodynamics for three-dimensional linearized
  {Grad}'s equations.
\newblock {\em Phys. Rev. E}, 76:022201, 2007.

\bibitem[CKO09]{colan09}
M.\ Colangeli, M.\ Kr\"{o}ger, and H.~C.\ \"{O}ttinger.
\newblock Boltzmann equation and hydrodynamic fluctuations.
\newblock {\em Phys. Rev. E}, 80:051202, 2009.

\bibitem[CM22]{CM22}
M.~Colangeli and A.~Muntean.
\newblock Reduced markovian descriptions of brownian dynamics: Toward an exact
  theory.
\newblock {\em Frontiers in Physics}, 10, 2022.

\bibitem[CMW11]{Col11}
M.~Colangeli, C.~Maes, and B.~Wynants.
\newblock A meaningful expansion around detailed balance.
\newblock {\em Journal of Physics A: Mathematical and Theoretical},
  44(9):095001, feb 2011.

\bibitem[CTDN20]{Checkroun1}
M.~D. Checkroun, A.~Tantet, H.~A. Dijkstra, and J.~D. Neelin.
\newblock Ruelle-{Pollicott} {Resonances} of {Stochastic} {Systems} in
  {Reduced} {State} {Space}. {Part} {I}: {Theory}.
\newblock {\em Journal of Statistical Physics}, 179:1366--1402, 2020.

\bibitem[DLP{\etalchar{+}}18]{Duong2018}
M.~H. Duong, A.~Lamacz, M.~A. Peletier, A.~Schlichting, and U.~Sharma.
\newblock Quantification of coarse-graining error in {L}angevin and overdamped
  {L}angevin dynamics.
\newblock {\em Nonlinearity}, 31(10):4517--4566, aug 2018.

\bibitem[DLPS17]{duong2017variational}
M.H. Duong, A.~Lamacz, M.~A. Peletier, and U.~Sharma.
\newblock Variational approach to coarse-graining of generalized gradient
  flows.
\newblock {\em Calc. Var. Partial Differ. Equ.}, 56:1--65, 2017.

\bibitem[DN23]{duong2023asymptotic}
M.~H. Duong and H.~D. Nguyen.
\newblock Asymptotic analysis for the generalized langevin equation with
  singular potentials.
\newblock {\em arXiv preprint arXiv:2305.03637}, 2023.

\bibitem[DS22]{duong2022accurate}
M.~H. Duong and X.~Shang.
\newblock {Accurate and robust splitting methods for the generalized Langevin
  equation with a positive Prony series memory kernel}.
\newblock {\em J. Comput. Phys.}, 464:111332, 2022.

\bibitem[GK05]{GorKar05}
A.~N.\ Gorban and I.~V.\ Karlin.
\newblock {\em Invariant Manifolds for Physical and Chemical Kinetics}, volume
  660 of {\em Lect. Notes Phys.}
\newblock Springer-Verlag, Berlin, 2005.

\bibitem[GK13]{GorKar}
A.~N.\ Gorban and I.~V.\ Karlin.
\newblock Hilbert's $6$th problem: {E}xact and approximate manifolds for
  kinetic equations.
\newblock {\em Bulletin of the American Mathematical Society}, 51:187--246,
  2013.

\bibitem[GKK{\etalchar{+}}06]{gorban2006model}
A.~N Gorban, N.~K. Kazantzis, I.~G. Kevrekidis, H.~C. {\"O}ttinger, and
  C.~Theodoropoulos.
\newblock {\em Model reduction and coarse-graining approaches for multiscale
  phenomena}.
\newblock Springer, 2006.

\bibitem[GKS04]{givon2004extracting}
D.~Givon, R.~Kupferman, and A.~Stuart.
\newblock Extracting macroscopic dynamics: model problems and algorithms.
\newblock {\em Nonlinearity}, 17(6):R55, 2004.

\bibitem[GKZ04]{Gor04}
A.~N. Gorban, I.~V. Karlin, and A.~Yu. Zinovyev.
\newblock Constructive methods of invariant manifolds for kinetic problems.
\newblock {\em Physics Reports}, 396(4):197--403, 2004.

\bibitem[GLCG21]{Ghil}
M.~S.\ Guti\'{e}rrez, V.\ Lucarini, M.~D.\ Chekroun, and M.\ Ghil.
\newblock Reduced-order models for coupled dynamical systems: Data-driven
  methods and the {K}oopman operator.
\newblock {\em Chaos}, 31:053116, 2021.

\bibitem[Has76]{hasselmann1976stochastic}
K.~Hasselmann.
\newblock Stochastic climate models {Part I. Theory}.
\newblock {\em Tellus}, 28(6):473--485, 1976.

\bibitem[HJ94]{haunggi1994colored}
P.~H{\"a}nggi and P.~Jung.
\newblock Colored noise in dynamical systems.
\newblock {\em Advances in chemical physics}, 89:239--326, 1994.

\bibitem[HNS20]{Hartmann2020}
C.~Hartmann, L.~Neureither, and U.~Sharma.
\newblock Coarse graining of nonreversible stochastic differential equations:
  Quantitative results and connections to averaging.
\newblock {\em SIAM Journal on Mathematical Analysis}, 52(3):2689--2733, 2020.

\bibitem[HSZ16]{hartmann}
C.~Hartmann, C.~Sch{\"u}tte, and W.~Zhang.
\newblock Model reduction algorithms for optimal control and importance
  sampling of diffusions.
\newblock {\em Nonlinearity}, 29(8):2298--2326, 2016.

\bibitem[KCK08]{colan08}
I.~V.\ Karlin, M.\ Colangeli, and M.\ Kr\"{o}ger.
\newblock Exact linear hydrodynamics from the {Boltzmann} equation.
\newblock {\em Phys. Rev. Lett.}, 100:214503, 2008.

\bibitem[KG02]{Kar02}
I.~V.\ Karlin and A.~N.\ Gorban.
\newblock Hydrodynamics from {Grad}'s equations: {W}hat can we learn from exact
  solutions?
\newblock {\em Annalen der Physik}, 11:783--833, 2002.

\bibitem[Kub66]{Kubo66}
R.\ Kubo.
\newblock The fluctuation-dissipation theorem.
\newblock {\em Rep. Prog. Phys.}, 29:255, 1966.

\bibitem[Kup04]{kupferman2004fractional}
Raz Kupferman.
\newblock {Fractional kinetics in Kac--Zwanzig heat bath models}.
\newblock {\em J. Stat. Phys.}, 114(1):291--326, 2004.

\bibitem[LC12]{LucCol12}
V.~Lucarini and M.~Colangeli.
\newblock Beyond the linear fluctuation-dissipation theorem: the role of
  causality.
\newblock {\em Journal of Statistical Mechanics: Theory and Experiment},
  2012(05):P05013, may 2012.

\bibitem[LC23]{lucarini2023theoretical}
V.~Lucarini and M.~D. Chekroun.
\newblock Theoretical tools for understanding the climate crisis from
  hasselmann’s programme and beyond.
\newblock {\em Nature Reviews Physics}, 5(12):744--765, 2023.

\bibitem[LL10]{Legoll2010}
F.~Legoll and T.~Leli{\`e}vre.
\newblock Effective dynamics using conditional expectations.
\newblock {\em Nonlinearity}, 23(9):2131, 2010.

\bibitem[LLS19]{Legoll2019}
F.~Legoll, T.~Leli{\`{e}}vre, and U.~Sharma.
\newblock Effective dynamics for non-reversible stochastic differential
  equations: a quantitative study.
\newblock {\em Nonlinearity}, 32(12):4779--4816, oct 2019.

\bibitem[LS22]{leimkuhler2022efficient}
B.~Leimkuhler and M.~Sachs.
\newblock Efficient numerical algorithms for the generalized langevin equation.
\newblock {\em SIAM Journal on Scientific Computing}, 44(1):A364--A388, 2022.

\bibitem[LVE14]{Lu2014}
J.~Lu and E.~Vanden-Eijnden.
\newblock Exact dynamical coarse-graining without time-scale separation.
\newblock {\em The Journal of Chemical Physics}, 141(4):044109, 2014.

\bibitem[LZ19]{Lelievre2019}
T.~Leli\`{e}vre and W.~Zhang.
\newblock Pathwise estimates for effective dynamics: {T}he case of nonlinear
  vectorial reaction coordinates.
\newblock {\em Multiscale Modeling \& Simulation}, 17(3):1019--1051, 2019.

\bibitem[MPRV08]{Vulp}
U.~B.~M.\ Marconi, A.\ Puglisi, L.\ Rondoni, and A.\ Vulpiani.
\newblock Fluctuation-{Dissipation}: Response {Theory} in {Statistical}
  {Physics}.
\newblock {\em Phys. Rep.}, 461:111--195, 2008.

\bibitem[Ngu18]{nguyen2018small}
H.~D. Nguyen.
\newblock {The small-mass limit and white-noise limit of an infinite
  dimensional generalized Langevin equation}.
\newblock {\em J. Stat. Phys.}, 173(2):411--437, 2018.

\bibitem[OP11]{ottobre2011asymptotic}
M.~Ottobre and G.~A. Pavliotis.
\newblock {Asymptotic analysis for the generalized Langevin equation}.
\newblock {\em Nonlinearity}, 24(5):1629, 2011.

\bibitem[{\"O}tt05]{Ott05}
H.~C. {\"O}ttinger.
\newblock {\em Beyond {Equilibrium} {Thermodynamics}}.
\newblock John {Wiley \& {Sons}}, 2005.

\bibitem[Pav14]{Pavl}
G.~A.\ Pavliotis.
\newblock {\em Stochastic Processes and Applications. Diffusion Processes, the
  {Fokker}-{Planck} and {Langevin} Equations}.
\newblock Springer-Verlag, 2014.

\bibitem[PS08]{pavliotis2008multiscale}
G.~Pavliotis and A.~Stuart.
\newblock {\em Multiscale methods: averaging and homogenization}.
\newblock Springer Science \& Business Media, 2008.

\bibitem[Ris96]{Risken}
H.\ Risken.
\newblock {\em The {Fokker}-{Planck} Equation}.
\newblock Springer Verlag, Berlin, 1996.

\bibitem[Rob15]{TonyRoberts}
A.~J. Roberts.
\newblock {\em Model {E}mergent {D}ynamics in {C}omplex {S}ystems}.
\newblock SIAM, 2015.

\bibitem[RVC22]{Rupe}
A.~Rupe, V.~V. Vesselinov, and J.~P. Crutchfield.
\newblock Nonequilibrium statistical mechanics and optimal prediction of
  partially-observed complex systems.
\newblock {\em New Journal of Physics}, 24:103033, 2022.

\bibitem[Sch09]{schuss2009theory}
Z.~Schuss.
\newblock {\em Theory and applications of stochastic processes: an analytical
  approach}, volume 170.
\newblock Springer Science \& Business Media, 2009.

\bibitem[SvdGT17]{Snowden}
T.~J. Snowden, P.~H. van~der Graaf, and M.~J. Tindall.
\newblock Methods of {M}odel {R}eduction for {L}arge-{S}cale {B}iological
  {S}ystems: {A} {S}urvey of {C}urrent {M}ethods and {T}rends.
\newblock {\em Bullettin of Mathematical Biology}, 79:1449--1486, 2017.

\bibitem[TCDN20]{Checkroun2}
A.~Tantet, M.~D. Checkroun, H.~A. Dijkstra, and J.~D. Neelin.
\newblock Ruelle-{Pollicott} {Resonances} of {Stochastic} {Systems} in
  {Reduced} {State} {Space}. {Part} {II}: {Stochastic} {Hopf} {Bifurcation}.
\newblock {\em Journal of Statistical Physics}, 179:1403--1448, 2020.

\bibitem[WG19a]{Wouters2019b}
J.~Wouters and G.~A. Gottwald.
\newblock Edgeworth expansions for slow-fast systems with finite time-scale
  separation.
\newblock {\em Proceedings of the Royal Society A: Mathematical, Physical and
  Engineering Sciences}, 475(2223):20180358, 2019.

\bibitem[WG19b]{Wouters2019}
J.~Wouters and G.~A. Gottwald.
\newblock Stochastic model reduction for slow-fast systems with moderate time
  scale separation.
\newblock {\em Multiscale Modeling \& Simulation}, 17(4):1172--1188, 2019.

\bibitem[ZHS16]{zhang2016effective}
W.~Zhang, C.~Hartmann, and C.~Sch{\"u}tte.
\newblock Effective dynamics along given reaction coordinates, and reaction
  rate theory.
\newblock {\em Faraday discussions}, 195:365--394, 2016.

\bibitem[Zwa01]{Zwanzig}
R.\ Zwanzig.
\newblock {\em Nonequilibrium {S}tatistical {M}echanics}.
\newblock Oxford University Press, 2001.

\end{thebibliography}
\end{document}